# Applications of the series expansion unknown functions method in nonlinear dynamics of microtubules

S. ZDRAVKOVIĆ AND S. ZEKOVIĆ

Institut za nuklearne nauke Vinča, Univerzitet u Beogradu, Laboratorija za atomsku fiziku (040), Poštanski fah 522, 11001 Beograd, Serbia

Microtubules (MTs) are the most important part of cytoskeleton. In this paper we deal with two basic nonlinear differential equations coming from the two known models describing nonlinear dynamics of MTs. These equations are solved using the series expansion unknown functions method (SEUFM). Trying to recognize the most general mathematical procedure for solving these equations the solutions are compared with those obtained earlier using the tangent hyperbolic function method (THFM) and the simplest equation method (SEM). In all these three approaches we express the solutions of these equations as series expansions. In the cases of THFM and SEM the functions existing in the series are known while SEUFM assumes unknown functions.

## 1. Introduction

Microtubules (MTs) form an important part of the cellular skeleton and represent a network for intracellular transport of motor proteins. Their structure is known [1-8] and we explain it very shortly. MT is a long holly cylinder consisting of usually 13 long structures called protofilaments (PFs) whose constituent parts are dimers. The dimers are electric dipoles and perform radial oscillations. A displacement of the dimer from its ground position can be expressed using either an angle $\varphi$ or a longitudinal coordinate $u$, representing a projection of the top of the dimer in the direction of PF [3]. Hence, we talk about a $\varphi$-model [9] and a $u$-model [10], respectively. Initial Hamiltonians, for these two models, in the nearest neighbour approximation, are [9,10]

$$H_{\varphi} = \sum_n \left[ \frac{I}{2} \dot{\varphi}_n^{\ 2} + \frac{k}{2} (\varphi_{n+1} - \varphi_n)^2 - pE \cos \varphi_n \right] \tag{1}$$

and

$$H_u = \sum_n \left[ \frac{m}{2} \dot{u}_n^{\ 2} + \frac{k}{2} (u_{n+1} - u_n)^2 - \frac{1}{2} A u_n^2 + \frac{1}{4} B u_n^4 - C u_n \right], \quad C = qE \,, \tag{2}$$

where the dot means a first derivative with respect to time, $m$ and $I$ are mass and moment of inertia of the single dimer, respectively, and $k$ stands for inter-dimer bonding interaction within the same PF. The integer $n$ determines the position of the dimer in the PF, $E > 0$ is the magnitude of the intrinsic electric field at the site *n*, as the dimer *n* exists in the electric field of all other dimers, $q > 0$ represents the excess charge within the dipole and $p > 0$ is electric dipole moment of the single dimer. Finally, *A* and *B* are positive constants that should be determined. We use the same notation for *k* in both equations for short.

Using continuum approximation we obtain crucial partial differential equations

$$I\frac{\partial^2\varphi}{\partial t^2} - kl^2\frac{\partial^2\varphi}{\partial x^2} + pE\varphi - \frac{pE}{6}\varphi^3 + \Gamma\frac{\partial\varphi}{\partial t} = 0 \tag{3}$$

and

$$m\frac{\partial^2 u}{\partial t^2} - kl^2\frac{\partial^2 u}{\partial x^2} - qE - Au + Bu^3 + \gamma\frac{\partial u}{\partial t} = 0, \tag{4}$$

where *x* and *t* represent space and time coordinates, respectively, while $l = 8\text{nm}$ is the dimer's length. The last terms in both equations are coming from viscosity momentum $-\Gamma\dot{\varphi}$ and viscosity force $-\gamma\dot{u}$ introduced in dynamical equations of motion [3,9,10].

We are looking for travelling wave solutions depending upon $x$ and $t$ only through a unified variable

$$\xi \equiv \kappa x - \omega t, \tag{5}$$

where $\kappa$ and $\omega$ are constants. Also, it is convenient to introduce dimensionless functions $\psi \equiv \psi(\xi)$ through

$$\varphi = \psi\sqrt{6}\,, \qquad u = \sqrt{\frac{A}{B}}\,\psi_u\,, \tag{6}$$

where, of course, the index $u$ has been introduced to distinguish the solutions for the two studied models. All this brings about the final ordinary differential equations (ODEs) for the both models

$$\alpha\frac{d^2\psi}{d\xi^2} - \rho\frac{d\psi}{d\xi} + \psi - \psi^3 = 0 \tag{7}$$

and

$$\alpha\frac{d^2\psi_u}{d\xi^2} - \rho\frac{d\psi_u}{d\xi} - \psi_u + \psi_u{}^3 - \sigma = 0\,. \tag{8}$$

We use the same notation for $\alpha$ and $\rho$ in both equations for short. Expressions for $\alpha$, $\rho > 0$ and $\sigma > 0$ are explained in Refs. [9,10]. It is important to keep in mind that $\rho$ in Eqs. (7) and (8) is proportional to the viscosity coefficients $\Gamma$ and $\gamma$, respectively, while the parameter $\sigma$ is proportional to the internal electric field strength *E*. We assume that $\rho$ and $\sigma$ are known and try to determine $\alpha$.

Equatios (7) and (8) will be solved in Sections 2, 3 and 4. A common procedure to solve ODE is to express its solution as a series expansion, like

$$\psi = A_0 + \sum_{k=1}^{N} A_k f^k \ . \tag{9}$$

If the function $f$ is known we plug Eq. (9) into the equation we want to solve, like Eqs. (7) and (8) in our case, and determine the coefficients $A_0$ and $A_k$. For example, if $f$ is a solution of Riccati equation the procedure is called the tangent hyperbolic function method (THFM) [11-14]. Possible variants would be extended THFM or modified extended THFM [15]. The function $f$ can also be one of Jacobian elliptic functions [16-18] and so on. It is very likely that the most general procedure is the simplest equation method (SEM) [19,20] and its simplified version called as modified simplest equation method (MSEM) [21]. Common for all these methods is a fact that the function $f$ is known. However, it is possible to use the series expansion over unknown functions [22]. In this paper we call this procedure as the series expansion unknown function method (SEUFM) as we want to stress the fact that these functions are unknown. We plug Eq. (9) into the one that we solve and try to obtain not only the parameters but the function $f$ as well. Intuitively, we expect that this method could be more general than any other as $f$ is not known in advance.

The purpose of this paper is to compare SEUFM with extended THFM and SEM. In fact, Eqs. (7) and (8) have already been solved using extended THFM [9] and SEM [10], respectively. One particular case regarding (7) was not studied in Ref. [9] and the appropriate derivation is explained in Section 2. Sections 3 and 4 are devoted to SEUFM. Finally, some concluding remarks are in Section 5.

## 2. Solutions of Eq. (7) using SEM

As was stated above, equation (7) describes nonlinear dynamics of MTs relying on the $\varphi$-model. It has been already solved using SEM [9]. Hence, we here very briefly repeat its solutions. Also, one particular case was not studied in [9] and it is explained here. A trial function is

$$\psi = A_0 + \sum_{k=1}^{N} \left( A_k \Phi^k + B_k \left( \frac{\Phi'}{\Phi} \right)^k \right), \tag{10}$$

where $A_0$, $A_k$ and $B_k$ are coefficients that should be determined and $\Phi'$ represents the first derivative. The function $\Phi$ is a known solution of a so-called the simplest equation which can be any nonlinear ordinary differential equation of lower order than Eq. (7) with the known general solution [19,20]. An example used in [9] is the well known Riccati equation

$$\Phi' + \Phi^2 - 2a\Phi - b = 0, \qquad a, b = \text{const}. \tag{11}$$

One can easily show that $N = 1$ for Eq. (7) and that a solution of Eq. (11) is

$$\Phi = a + \sqrt{a^2 + b}\,\tanh\left[\sqrt{a^2 + b}\left(\xi - \xi_0\right)\right], \tag{12}$$

where $\xi_0$ is an arbitrary constant of integration [9]. In what follows we assume $\xi_0 = 0$.

Equations (7) and (10) bring about a system of 7 equations. One of them is

$$\left(B_1 - A_1\right)\left[\left(B_1 - A_1\right)^2 - 2\alpha\right] = 0, \tag{13}$$

which determines the following two cases: $B_1 = A_1$ and $\alpha = \left(A_1 - B_1\right)^2/2$. The latter case yields to $A_1 = 2B_1$ and $a = 0$. The solutions, corresponding to these two cases, are

$$\psi_1(x,t) = \pm\frac{1}{2}\left[1 + \tanh y + \frac{1}{\cosh^2 y\,\left(d + \tanh y\right)}\right] \tag{14}$$

and

$$\psi_2(x,t) = \pm\frac{1}{2}\left[1 + \tanh\left(\frac{y}{2}\right) + \frac{1}{\sinh y}\right], \tag{15}$$

respectively, where

$$y = \frac{3}{4\rho}\xi, \qquad d = \frac{4\rho}{3}a. \tag{16}$$

If $b = 0$ and $A_1 = B_1$ then Eq. (7) has trivial solutions $\psi = 0$ and $\psi = \pm 1$ only [9]. For $a = 0$ and $b = 0$ the solution of Eq. (11) is $\Phi = 1/\xi$. This means that it does not make sense to assume $b = 0$ within the 2nd case, i.e. the one determined by $\alpha = (A_1 - B_1)^2/2$.

For $B_1 = 0$ SEM becomes extended THFM and the solution we are looking for is

$$\psi_3(x,t) = \pm\frac{1}{2}[1 + \tanh y]. \qquad (17)$$

We can see that $\psi_1$ for $d > 1$ and $\psi_3$ represent kink solitons. It was shown in Ref. [9] that the solutions $\psi_1$ and $\psi_3$ are equivalent, at least from the physics point of view, as

$$\psi_3(y + \delta) = \psi_1(y), \qquad \tanh\delta = 1/d. \qquad (18)$$

Hence, these two functions equally describe nonlinear dynamics of MTs, while $\psi_2$, diverging for $\xi = 0$, does not have physical meaning.

A patient reader may have noticed that the solution $\psi_3$ was obtained for $B_1 = 0$. However, the case $A_1 = 0$ has not been mentioned yet. This case was not studied in Ref. [9] either. Hence, in what follows, we explain the solutions of Eq. (7) for $A_1 = 0$. When we plug Eq. (10) into (7) we obtain the expression

$$K_0 + K_1\Phi + K_1'\Phi^{-1} + K_2\Phi^2 + K_2'\Phi^{-2} + K_3\Phi^3 + K_3'\Phi^{-3} = 0, \qquad (19)$$

which is satisfied if all the coefficients $K_i$ and $K_i'$ are simultaneously equal to zero. This brings about a system of seven equations. When we introduce $A_1 = 0$ into the system we obtain

$$\alpha = \frac{B_1^2}{2}. \tag{20}$$

Further inspection of the system shows that there are two possibilities. These are: $b \neq 0$ and $b = 0$. For $b \neq 0$ the system gives

$$\rho = 0, \quad A_0 = -aB_1, \quad a = 0, \quad B_1^2 = -\frac{1}{2b}. \tag{21}$$

Obviously, this is valid only for negative $b$. Also, this case does not have real physical meaning as the parameter $\rho$ is proportional to viscosity and, therefore, cannot be equal to zero. Of course, we may talk of the case when viscosity in neglected. Using Eqs. (10), (12) and (21), as well as the formula $\tanh(i\beta) = i \tan\beta$, where $i$ is an imaginary unit, we obtain the final solution

$$\psi_4 = \pm\frac{\sqrt{2}}{\sin\left(2\sqrt{-b}\xi\right)}, \quad b < 0. \tag{22}$$

Of course, this solution diverges for $\sqrt{-b}\xi = k\pi$.

For $b = 0$ the aforementioned system yields to the following four pairs of $a$ and $B_1$:

$$a = \pm \frac{3}{4\rho}, \quad B_1 = \pm \frac{2\rho}{3}, \tag{23}$$

as well as

$$A_0 = -aB_1 - \frac{\rho}{3B_1}. \tag{24}$$

All this brings about the solution $\psi_3$.

### 3. Solutions of Eq. (7) using SEUFM

In what follows, we solve Eq. (7) using SEUFM. The trial function is [22]

$$\psi = A_0 + \sum_{k=1}^{N} A_k \left( \frac{\Phi'}{\Phi} \right)^k, \quad \Phi' \neq 0, \tag{25}$$

where $N = 1$ for this case, $A_0$ and $A_1$ are coefficients that should be determined and, as was pointed out above, $\Phi$ is the unknown function. Therefore, according to Eqs. (7) and (25), we should obtain not only $A_0$ and $A_1$ but the function $\Phi$ as well. In addition, the parameter $\alpha$ will be also determined. Hence, when we plug Eq. (25) into (7) we obtain the coefficients for $\Phi^{-3}$, $\Phi^{-2}$, $\Phi^{-1}$ and $\Phi^{0}=1$ that should simultaneously be equal to zero. The first one, corresponding to $\Phi^{-3}$, gives

$$\alpha=\frac{A_1^{\,2}}{2} \tag{26}$$

and the remaining three, together with Eq. (26), yield to

$$-3A_1^{\,2}\Phi''+2(\rho-3A_0A_1)\Phi'=0\,, \tag{27}$$

$$A_1^{\,2}\Phi'''-2\rho\,\Phi''+2\left(1-3A_0^{\,2}\right)\Phi'=0 \tag{28}$$

and

$$A_0-A_0^{\,3}=0\,. \tag{29}$$

Equation (27) can be easily integrated and we obtain

$$\Phi=\frac{K_1}{\beta}e^{\beta\xi}+K_2\,,\qquad \beta=\frac{\rho-3A_0A_1}{3\alpha}\,, \tag{30}$$

where $K_1$ and $K_2$ are constants of integration. When we plug Eq. (30) into (28) we obtain a crucial equation

$$9\left(A_0^{\,2}-1\right)A_1^{\,2}-6\rho A_0 A_1+4\rho^2=0\,. \tag{31}$$

Obviously, possible values for $A_0$ are

$$A_0=0\,,\qquad A_0=\pm 1 \tag{32}$$

and all these values give

$$A_1=\pm\frac{2\rho}{3}\,,\qquad\qquad \alpha=\frac{2\rho^2}{9}\,, \tag{33}$$

which can be seen from Eqs. (26), (31) and (32). Notice that, for $A_0=\pm 1$, the parameters $A_0$ and $A_1$ have the same sign, as Eq. (31) yields to $A_0 A_1=2\rho/3>0$.

Therefore, we know $A_0$, $A_1$, $\alpha$ and $\Phi$ and, according to Eqs. (25), (30), (32) and (33), we finally obtain

$$\psi(\xi)=\pm\frac{e^{\frac{3}{2\rho}\xi}}{e^{\frac{3}{2\rho}\xi}+C}\,, \tag{34}$$

where the constant $C$ was introduced through

$$K_2 = K_1 \frac{2\rho}{3} C\,. \tag{35}$$

One can easily study the cases $C = \pm 1$ using formulae

$$\frac{e^x}{e^x + 1} = \frac{1}{2}\left[1 + \tanh\left(\frac{x}{2}\right)\right], \qquad \frac{e^x}{e^x - 1} = \frac{1}{2}\left[1 + \coth\left(\frac{x}{2}\right)\right]. \tag{36}$$

We see that $\psi = \psi_3$ for $C = 1$. Other positive values for $C$ give different solutions but, from physics point of view, all of them are equivalent. Namely, the constant $C$ represents only a shift. From

$$\frac{e^{\lambda\xi}}{e^{\lambda\xi} + C} = \frac{e^{\lambda(\xi+\delta)}}{e^{\lambda(\xi+\delta)} + 1} \tag{37}$$

we see that the shift is

$$\delta = -\frac{\ln C}{\lambda}\,. \tag{38}$$

For $C = -1$ the general solution (34) brings about the function

$$\psi_5(x,t) = \pm\frac{1}{2}\left[1 + \coth(y)\right] \tag{39}$$

instead of $\psi_3$, shown above, where $y$ is given by Eq. (16). Of course, for different but negative values of $C$ the graphs are parallel and Eq. (38) holds again except that $C$ should be replaced by $|C|$.

Therefore, SEUFM yields the same solution as SEM, having physical sense. This is the solution $\psi_3$. However, this procedure does not bring about the diverging solutions $\psi_2$ and $\psi_4$ but yields to infinitely many diverging solutions (34) corresponding to the negative $C$. One such example is the function $\psi_5$. On the other hand, the solution given by Eq. (34), corresponding to the positive $C$, is more general than $\psi_3$, which is an advantage of the present method. Also, an advantage of SEUFM is its simplicity. It is obvious that the series expansion (25) is simpler than the one given by Eq. (10). This means that there are not so many constants that should be determined in Eq. (25) as there are in Eq. (10).

Before we proceed we want to give a short comment on Eq. (34). This solution is common for both $A_0 = 0$ and $A_0 = \pm 1$. Suppose that the constants $C$ are $C_0$ for $A_0 = 0$ and $C_1$ for $A_0 = \pm 1$. These two lines, corresponding to the constants $C_0$ and $C_1$, match for $C_0 C_1 = 1$.

## 4. Solutions of Eq. (8) using SEUFM

When we plug Eq. (25) into (8) and study the coefficients for $\Phi^{-3}$, $\Phi^{-2}$, $\Phi^{-1}$ and $\Phi^0 = 1$ we obtain the following four equations

$$\alpha = -\frac{A_1^{\ 2}}{2}, \tag{40}$$

$$3A_1^{\ 2}\Phi'' + 2(\rho + 3A_0 A_1)\Phi' = 0, \tag{41}$$

$$A_1^{\ 2}\Phi''' + 2\rho\Phi'' + 2(1 - 3A_0^{\ 2})\Phi' = 0 \tag{42}$$

and

$$A_0^{\ 3} - A_0 - \sigma = 0. \tag{43}$$

These equations correspond to Eqs. (26)-(29) in Section 3. Notice the negative $\alpha$ in Eq. (40). The meaning of these signs is explained in Refs. [9,10].

Solutions of Eq. (43) are [10]

$$\left.\begin{aligned} A_{01} &= \frac{2}{\sqrt{3}}\cos F > 0, \\ A_{02} &= \frac{1}{\sqrt{3}}\left(-\cos F + \sqrt{3}\sin F\right) < 0, \\ A_{03} &= -\frac{1}{\sqrt{3}}\left(\cos F + \sqrt{3}\sin F\right) < 0 \end{aligned}\right\} \tag{44}$$

where

$$F = \frac{1}{3}\arccos\left(\frac{\sigma}{\sigma_0}\right), \qquad \sigma_0 = \frac{2}{3\sqrt{3}}. \tag{45}$$

We can easily check that

$$A_{01} = 1, \quad A_{02} = 0, \quad A_{03} = -1 \quad \text{for} \quad \sigma = 0, \tag{46}$$

and

$$A_{01} = \frac{2}{\sqrt{3}}, \quad A_{02} = A_{03} = -\frac{1}{\sqrt{3}}, \quad \text{for} \quad \sigma = \sigma_0. \tag{47}$$

Notice that $\sigma = 0$ does not have physical meaning as $\sigma$ is proportional to internal electric field strength, which is not zero. However, in this paper, we are looking for all possible solutions of Eq. (8). Also, the solutions corresponding to $\sigma = 0$ represent a limit when $\sigma$, i.e. electric field strength, is very small.

Instead of Eq. (31) we now obtain

$$9\left(A_0^{\,2} - 1\right)A_1^{\,2} + 6\rho A_0 A_1 + 4\rho^2 = 0, \tag{48}$$

while Eq. (30) becomes

$$\Phi = \frac{K_1}{\beta} e^{\beta\xi} + K_2, \qquad \beta = \frac{\rho + 3A_0 A_1}{3\alpha}. \tag{49}$$

Of course, $A_0$ stands for $A_{0i}$, $i = 1, 2, 3$, and, consequently, there are three possible values for $A_1$, which will be denoted as $A_{11}$, $A_{12}$ and $A_{13}$ in what follows.

Following the procedure explained in the previous sections we obtain, for $A_0 = 0$, the final solution

$$\psi_{u0}(\xi) = \pm \frac{e^{-\frac{3}{2\rho}\xi}}{e^{-\frac{3}{2\rho}\xi} + C}, \tag{50}$$

where, of course, $C$ is an arbitrary constant of integration. Equation (50), for $C = 1$ and $C = -1$, respectively gives

$$\psi_{u01} = \pm \frac{1}{2}\left[1 - \tanh y\right], \qquad \psi_{u02} = \pm \frac{1}{2}\left[1 - \coth y\right], \tag{51}$$

where $y$ is given by Eq. (16).

For $A_0 = \pm 1$ we obtain the same solution, i.e. the one given by Eq. (50). Strictly speaking, as was explained in the previous section, the solutions for $A_0 = 0$ and $A_0 = \pm 1$ are equal if the product of their constants is 1. Of course, $C$ is the arbitrary constant and we can say that these solutions are equal. Notice different signs in Eqs. (17) and (51). These two solutions are kink and anti-kink solitons but, from physics point of view, they are equal.

Let us study general solutions, which means that we assume $A_0 \neq 0$ and $A_0 \neq \pm 1$. From Eq. (48) we obtain

$$A_1 = \frac{\rho A_0 \pm \rho\sqrt{4 - 3A_0^{\ 2}}}{3\left(1 - A_0^{\ 2}\right)}. \qquad (52)$$

According to Eq. (44) the expressions $\sqrt{4-3A_0^{\ 2}}$ can be simplified as

$$\left.\begin{aligned} \sqrt{4-3A_{01}^{\ 2}} &= 2\sin F, \\ \sqrt{4-3A_{02}^{\ 2}} &= \sqrt{3}\cos F + \sin F, \\ \sqrt{4-3A_{03}^{\ 2}} &= \sqrt{3}\cos F - \sin F > 0. \end{aligned}\right\}. \qquad (53)$$

Due to the two possible signs in Eq. (52) there are six values for the parameter $A_1$. These are

$$\left.\begin{aligned} A_{11}^{\ +} &= \frac{2\rho}{3A_{02}}, \qquad & A_{11}^{\ -} &= \frac{2\rho}{3A_{03}}, \\ A_{12}^{\ +} &= \frac{2\rho}{3A_{01}}, \qquad & A_{12}^{\ -} &= \frac{2\rho}{3A_{03}}, \\ A_{13}^{\ +} &= \frac{2\rho}{3A_{01}}, \qquad & A_{13}^{\ -} &= \frac{2\rho}{3A_{02}} \end{aligned}\right\} \qquad (54)$$

Also, Eqs. (40), (49) and (54) yield to all possible values for $\beta$ that are

$$
\left.\begin{aligned}
\beta_1^{+} &= -\frac{3}{2\rho}A_{02}(2A_{01}+A_{02}), & \beta_1^{-} &= -\frac{3}{2\rho}A_{03}(2A_{01}+A_{03}) \\
\beta_2^{+} &= -\frac{3}{2\rho}A_{01}(2A_{02}+A_{01}), & \beta_2^{-} &= -\frac{3}{2\rho}A_{03}(2A_{02}+A_{03}) \\
\beta_3^{+} &= -\frac{3}{2\rho}A_{01}(2A_{03}+A_{01}), & \beta_3^{-} &= -\frac{3}{2\rho}A_{02}(2A_{03}+A_{02})
\end{aligned}\right\} \tag{55}
$$

Therefore, we can write the final solutions. According to Eqs. (25) and (49) they are

$$
\psi_u = A_0 + A_1\beta \frac{e^{\beta\xi}}{e^{\beta\xi}+C}, \tag{56}
$$

where $C$ was introduced through

$$
K_2 = \frac{K_1}{\beta}C, \tag{57}
$$

while $A_0$, $A_1$ and $\beta$ stand for $A_{0i}$, $A_{1i}^{\pm}$ and $\beta_i^{\pm}$, $i = 1,2,3$. Obviously, Eq. (56) comprises six solutions that can be written as $\psi_i^{\pm}$, determined by the six sets of the corresponding parameters. However, a careful inspection of all the solutions shows that there are only three pairs of them, i.e. three solutions only. These solutions can be compared with those obtained using THFM [10]. If the solutions in Ref. [10] are denoted as $\psi_1$, $\psi_2$ and $\psi_3$, then the equality of the solutions are given by

$$
\psi_1^{-} = \psi_2^{-} = \psi_1, \qquad \psi_1^{+} = \psi_3^{-} = \psi_2, \qquad \psi_2^{+} = \psi_3^{+} = \psi_3. \tag{58}
$$

These solutions are given in Fig. 1 for $\sigma = 0.8\sigma_0$ and $C = 1$, obviously representing kink and antikink solitons. Of course, as explained above, for different values of the positive $C$ the graphs will be parallel those in Fig. 1. For negative $C$ the solutions diverge.

Equation (44) represents the solutions of Eq. (43) for $\sigma < \sigma_0$. If $\sigma > \sigma_0$ then there is only one real solution. This solution of Eq. (43) is [10]

$$A_0 = -\frac{2}{\sqrt{3}\sin(2\theta)}, \tag{59}$$

where

$$\tan\theta = -\sqrt[3]{\tan\left[0.5\arcsin\left(\frac{\sigma_0}{\sigma}\right)\right]}, \qquad \sigma > \sigma_0. \tag{60}$$

This means that the expression $4 - 3A_0^{\,2}$, existing in Eq. (52), is negative. Consequently, $A_1$ is not real and the solution of Eq. (8), for $\sigma > \sigma_0$, does not exist.

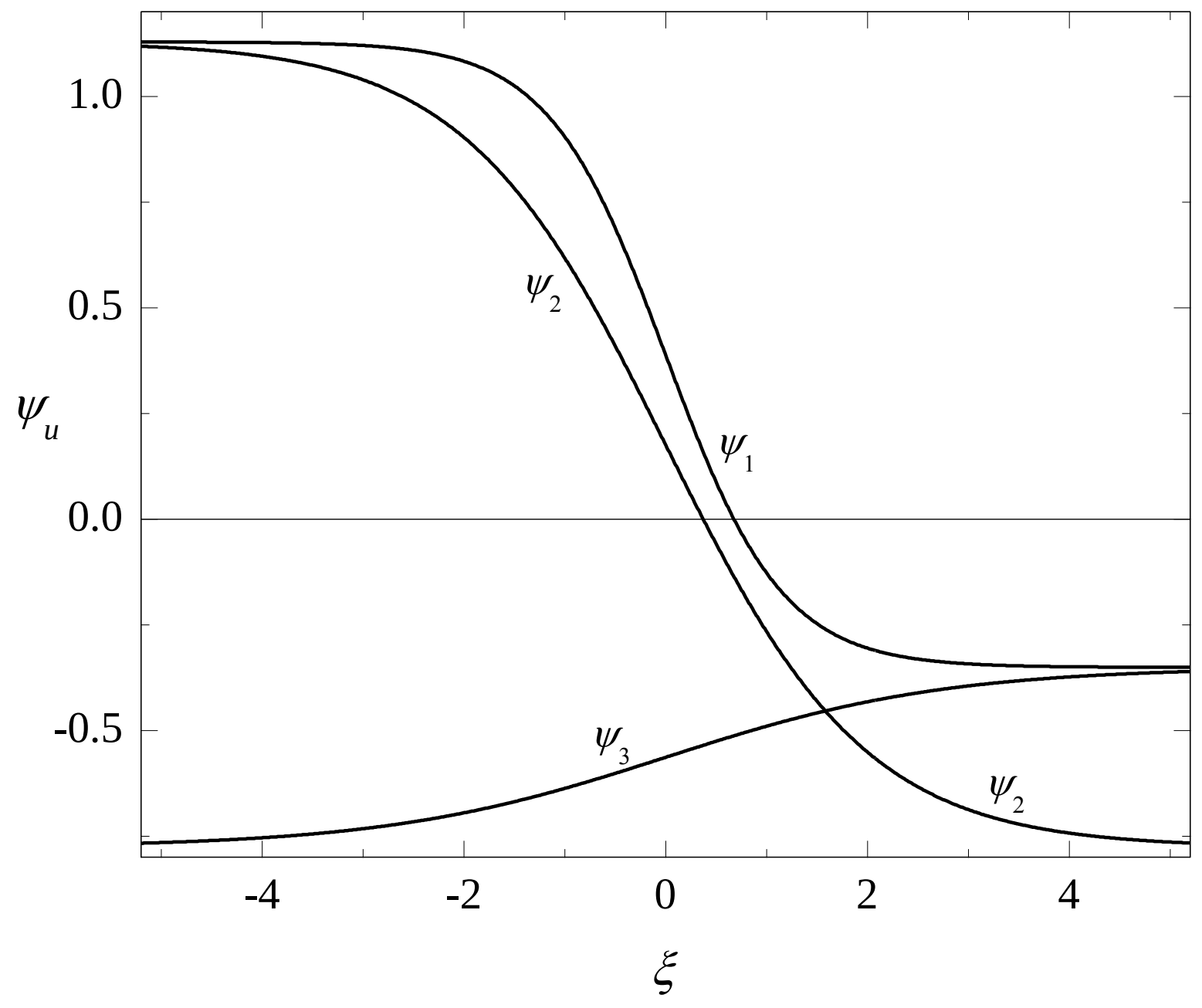

Fig. 1 - Functions $\psi(\xi)$ for $\sigma = 0.8\sigma_0$ and $C = 1$

As a conclusion we can say that the solutions obtained using THFM [10] represent special cases of those obtained using SEUFM. Namely, various values of the parameter $C$ give infinitely many solutions. The case existing in Ref. [10] corresponds to the special case when $C = 1$. As was explained above, all of them have equal physical meaning. For negative $C$ we obtain diverging solutions where tangent hyperbolic should be replaced with cotangent hyperbolic, which was also obtained in Ref. [10]. Also, THFM yields toward the diverging solutions where hyperbolic tangent and cotangent should be replaced by tangent and cotangent functions. These two cases were not obtained using SEUFM. Also, SEUFM does not bring about any solution for $\sigma > \sigma_0$, while THFM yields to the diverging one.

The readers may have noticed that, for the *u*-model, we compared our results obtained using SEUFM with those obtained using THFM. However, regarding the $\varphi$-model, we compared our new results, obtained using SEUFM, with the more general SEM. This is so because Eq. (8) has not been solved using SEM yet. However, our preliminary results show that the conclusions regarding both models are equal. This means that the most important solutions, having physical sense, shown in Fig. 1, were also obtained using SEM. Like above, the procedure explained in this paper yield to infinitely many soluitons. On the other hand, SEM brings about two additional diverging solutions that do not describe nonlinear dynamics of MTs.

## 5. Concluding remarks

In this paper we solved two basic equations describing MT dynamics using SEUFM. A key point regarding this procedure is the fact that the function used for the series expansion is not known. We wanted to figure it out if this fact could yield to some new solutions. Our conclusion, valid for both equations, is that, using SEUFM, we obtained infinitely many solutions having physical sense instead of only one. However, this is probably of interest for mathematics but, regarding nonlinear MT dynamics, this is not very important as all of them have equal physical meaning. Also, the very good side of this method is its simplicity. In addition, we did not obtain two solutions obtained using

SEM, which is also not important for physics of MTs as these solutions do not have physical sense.

As a general conclusion we stress our strong impression that the best that could and should be done is using more than one mathematical procedure. We do believe that SEM and SEUFM are the best choice.

Of course, all these conclusions are coming from MT studies only. It is important to figure it out if these conclusions are valid in general, i.e. for any differential equation, which should be future research

## Acknowledgement

This research was supported by funds from Serbian Ministry of Education, Sciences and Technological Development, grant III45010.